\documentclass{PoS}
\usepackage[compress,numbers]{natbib}
\def \be {\begin{equation}}
\def \ee {\end{equation}}
\def \ba {\begin{eqnarray}}
\def \ea {\end{eqnarray}}

\newcommand{\calC}{{\mathcal C}}

\newcommand{\calO}{{\mathcal O}}

\newcommand{\calZ}{{\mathcal Z}}
\title{Review on Hadron Spectroscopy}

\ShortTitle{Review on Hadron Spectroscopy}

\author{\speaker{Chuan Liu}\\
        School of Physics and Center for Hight Energy Physics\\
        Peking University, Beijing 100871, China\\
        \\
        Collaborative Innovation Center of Quantum Matter\\
        Peking University, Beijing 100871, China\\
        E-mail: \email{liuchuan@pku.edu.cn}\\}

\abstract{I review some of the lattice results on spectroscopy and resonances in the past years. For the conventional hadron spectrum computations, focus has been put on the isospin breaking effects, QED effects, and simulations near the physical pion mass point. I then go through several single-channel scattering studies within L\"uscher formalism, a method that has matured over the past few years. The topics cover light mesons and also the charmed mesons, with the latter case intimately related to the recently discovered exotic $XYZ$ particles. Other possible related formalisms that are available on the market are also discussed.}

\FullConference{34th annual International Symposium on Lattice Field Theory\\
		24-30 July 2016\\
		University of Southampton, UK}

\begin{document}

\section{Introduction}

 Spectroscopy computations have always been an important ingredient and a benchmark
 in lattice chromodynamics (lattice QCD) since
 its very early ages. In recent years, owing to the development in both theoretical
 understandings (physically and algorithmically)
 and computer hardware, lattice QCD computations have gradually made the
 crossover into the precision era, see e.g. the latest compilation of the
 FLAG Working Group~\cite{Aoki:2016frl,Tarantino:2012mq} for more details.
 In the light hadron spectrum sector, for example, lattice computations have been
 performed by various groups using full dynamical quarks with different number of
 dynamical flavors and different fermion realizations
 and the final results are very encouraging, see e.g. Ref.~\cite{Durr:2008zz}.
 For hadrons involving heavy quarks, precise lattice computations
 also exist for the charmonia below the open charm threshold, see e.g. Ref~\cite{Donald:2012ga},
 and for the single charmed mesons~\cite{Dowdall:2012ab}.

 To accomplish these precise lattice computations, one has to control a number of systematic effects.
 These include finite lattice spacing errors (lattice artifacts), heavier than physical quark masses
 (the chiral limit, or more precisely, the physical quark mass limit),
 finite volume effects, isospin breaking errors and QED effects. Let us first briefly
 discuss these issues in the following.

 To control the finite lattice spacing errors, improved actions and/or finer lattices have been
 utilized. As for the chiral limit, people have also been able to simulate right on the physical point
 or close enough to the physical quark mass values so as to make use of variants of Chiral Perturbation theory to
 access the true physical values for interested quantities. In fact,
 lattice QCD has provided an extra probe than true experiments in Nature
 in the sense that one can conveniently study the quark mass dependence of any physical quantity at any
 value of the quark mass. This particular dependence usually provides much
 more information of the theory than only at a particular value, even though it
 is the true physical value. We will see an example of this in subsection~\ref{subsec:physical-point}.

 For the finite volume errors, since most hadrons are resonances instead of
 stable particles, scattering states are necessarily included together with
 the single-hadron states with the same quantum numbers.
 This is particularly important for unstable hadrons (that is, resonances) above some threshold
 under strong interaction. One of course then needs the connection between
 the discrete energy levels in a finite box and the scattering information
 parameterized by the $S$-matrix elements. The main theoretical framework, known as
 L\"uscher's formalism in the community, has been established for more than two decades.
 However, only in recent years, the applications of this formalism in
 real lattice computations have matured. As we will see in this review,
 now we are able to reproduce e.g. the rho resonance with rather good precision,
 a task that could only be dreamed of in the 90's.

 Another systematic effects comes from the isospin breaking and quantum electrodynamics (QED).
 In this problem, there is the famous mystery of neutron-proton mass splitting which
 is a subtle balance of the isospin breaking effects and the QED effects.
 In the past few years, two lattice groups, namely BMW and QCDSF/UKQCD,
 have studied this challenging problem and we will briefly discuss
 their results in subsection~\ref{subsec:pn-mass-splitting}.

 For the hadrons with heavy flavors, in recent years, partly due to the recent experimental progress in
 the so-called $XYZ$ particles, lattice computations have also played an active role.
 Various groups have studied these newly discovered structures in both the charm
 and bottom sectors. Although these lattice studies are still not systematic enough
 to really nail down the true nature of these exotic particles, these studies
 have certainly provided important non-perturbative information for these states.

 This short review is organized as follows. In Sec.~\ref{sec:theory},
 I will first recapitulate the basic theoretical formalisms that
 have been utilized in typical lattice spectrum computations. Apart from the L\"uscher's formalism
 that has been mentioned above, I will also discuss three other available formalisms on
 the market: the Hamiltonian Effective Field Theory method (HEFT),
 the HAL QCD  method and the Optical Potential method. Then, starting
 in Sec.~\ref{sec:real_stuff}, I will go over the developments in spectroscopy in recent two years or so:
 starting from the light hadron spectrum, then move on to the single-channel scattering of light
 mesons and charmed mesons, the latter topic is heated up recently due to
 experimental discoveries of new near-threshold structures.
 Some related developments involving bottom quarks will be mentioned as well.

 It should be noted though many topics that is related to spectroscopy is not discussed
 here. One important subject is the scattering involving baryons which is
 partly reviewed in Savage's talk on lattice nuclear physics~\cite{Savage:2016}.
 I will also refer to Wilson's topical talk,
 which on the lattice calculation using coupled-channel
 L\"uscher formalism~\cite{Wilson:2016}, for multi-channel
 lattice computations in hadron scattering.

% \myref{Need to modify this part completely!}

\section{Theoretical methods utilized in spectrum computations}
\label{sec:theory}

 In a typical lattice QCD spectrum computation, one targets a specific channel
 with designated quantum numbers. Then a collection of interpolating operators
 $\{\calO_\alpha, \alpha=1,2,\cdots,N_{op}\}$  is chosen which carry the same quantum numbers
 of interest. Using Monte Carlo simulations,
 the following correlation matrix is estimated numerically,
 \be
 \calC_{\alpha\beta}(t)=\left\langle\calO_\alpha(t)\calO^\dagger_\beta(0)\right\rangle
 \;,
 \ee
 where $\langle\cdots\rangle$ stands for the expectation value in the QCD vacuum
 which is usually achieved by using a sample of gauge field configurations.
 By solving the so-called generalized eigen-value problem (GEVP) for a
 judiciously chosen time-slice $t_0$:
 \be
 \label{eq:Ealphas}
 \calC(t)\cdot u_\alpha=\lambda_\alpha(t,t_0)\calC(t_0)\cdot u_\alpha
 \;,\;\;\lambda_\alpha(t,t_0)\simeq e^{-E_\alpha(t-t_0)}\;,
 \ee
 one obtains the generalized eigenvalues $\lambda_\alpha(t,t_0)$.
 These eigenvalues are related to the exact energy eigenvalues, $E_\alpha$,
 of the QCD Hamiltonian via $\lambda_\alpha(t,t_0)\simeq e^{-E_\alpha(t-t_0)}$.
 Therefore, through such a process, one obtains the exact energy eigenvalues
 of the system in a particular channel.
 It should be noted that, these operators share the same good quantum numbers
 respected by the QCD Hamiltonian, in other words, all operators that carry the same
 quantum numbers mix within QCD. This includes single hadron operators, two-hadron operators, etc.
 In fact, most resonances can be studied using this method as
 will be illustrated in subsection~\ref{subsec:pipi-scattering} and \ref{subsec:XYZ}.

 The next step in a spectroscopy calculation relies on
 how these energy eigenvalues, the $E_\alpha$'s, are treated.
 It is clear that, at least in principle, these eigenvalues are not hadron mass values themselves,
 although they will approximate the hadron masses if the hadron resonance being considered is narrow enough.
 For generic resonances, it is also known that they are related to the scattering matrix
 elements within the so-called L\"uscher's formalism, see e.g. Ref.~\cite{Luscher:1986pf,Luscher:1990ux}.
 The theoretical formalism brought forward by L\"uscher, first illustrated for single-channel
 scattering of two identical spinless bosons in center-of-mass frame,
 was later on generalized in many ways so as to deal with more complicated scattering processes of
 various situations of the two hadrons in single or even in multi-channel scattering~\cite{Rummukainen:1995vs,Beane:2003da,Beane:2003yx,Detmold:2004qn,Li:2003jn,Feng:2004ua,%
 Kim:2005gf,Bernard:2008ax,Bour:2011ef,Davoudi:2011md,Gockeler:2012yj,Briceno:2013lba,%
 He:2005ey,Liu:2005kr,Hansen:2012tf,Briceno:2012yi,Li:2012bi,%
 Guo:2012hv,Bernard:2010fp,Roca:2012rx,Li:2014wga}.
 Up to now the theoretical formalism capable of dealing with the most general two-particle
 to two-particle scattering (single or multi-channel)
 is available and one in principle can utilize this to relate $S$-matrix elements to the
 eigenvalues obtained from GEVP.
 For the single-channel case, the formalism is rather straightforward.
 The practical lattice computations have matured over the past few years and
 some examples will be discussed in the following sections.
 For the case of multi-channel scattering,
 the application of the formalism is more elaborate and complicated.
 One needs  some concrete parameterizations of the $S$-matrix to proceed.
 I refer to the contribution of Hadron Spectrum Collaboration (HSC)
 in these proceedings~\cite{Wilson:2016} for concrete examples.

 Let us briefly review the main ideas behind L\"uscher's formalism.
 In its simplest case, one considers two identical spinless bosons with mass $m$ interact
 via a short-ranged interaction with range parameterized by $R$.
 In the center of mass frame, the elastic scattering phase
 of the two particle is described by the scattering phases $\delta_l(k)$
 where $k$ is the scattering momentum of the particle in the center of mass frame
 while $l$ designates various partial waves.
 If the two particles are put inside a box of size $L\gg R$,
 then the infinite-volume scattering phase $\delta_0(k)$
  \footnote{This assumes that $s$-wave scattering dominates, neglecting the
 contributions of higher partial waves. This is true in the case of
 near-threshold scattering to be discussed in the following.}
 is related to the exact energy of the two-particle system inside the box, $E(L)$,
 via,
 \be
 \tan\delta_0(\bar{k})=\frac{\pi^{3/2}q}{\calZ_{00}(1;q^2)}\;,
 \ee
 where $q$ and $\bar{k}$ are related by $q=\bar{k}L/(2\pi)$ and
 $\bar{k}$ is further related to $E(L)$ via
 $2\sqrt{\bar{k}^2+m^2}=E(L)$.
 Thus, by measuring the values of $E(L)$, which
 are nothing but those $E_\alpha$'s obtained from the GEVP process in
 Eq.~(\ref{eq:Ealphas}), one can extract the scattering phase
 shift $\delta_0(\bar{k})$ at those energies.

 Although L\"uscher's formalism has been established for almost two decades,
 the real lattice simulations using the formalism have only matured
 in recent years, particularly in the single-channel scenario. In the
 multichannel case, the usage of the formalism tends to be
 rather involved and complicated, see e.g. Ref.~\cite{Wilson:2016}.
 It is therefore desirable to search for other theoretical formalisms.
 Up to now, three methods have been put forward: the Hamiltonian Effective Field Theory (HEFT)
 approach~\cite{Hall:2013qba,Liu:2015ktc},
 the HAL QCD method~\cite{Ishii:2006ec,HALQCD:2012aa} and
 the Optical Potential (OP) method~\cite{Agadjanov:2016mao}.
 These methods  will be briefly discussed below.

 \begin{figure}[htb]
\includegraphics[width=0.50\textwidth]{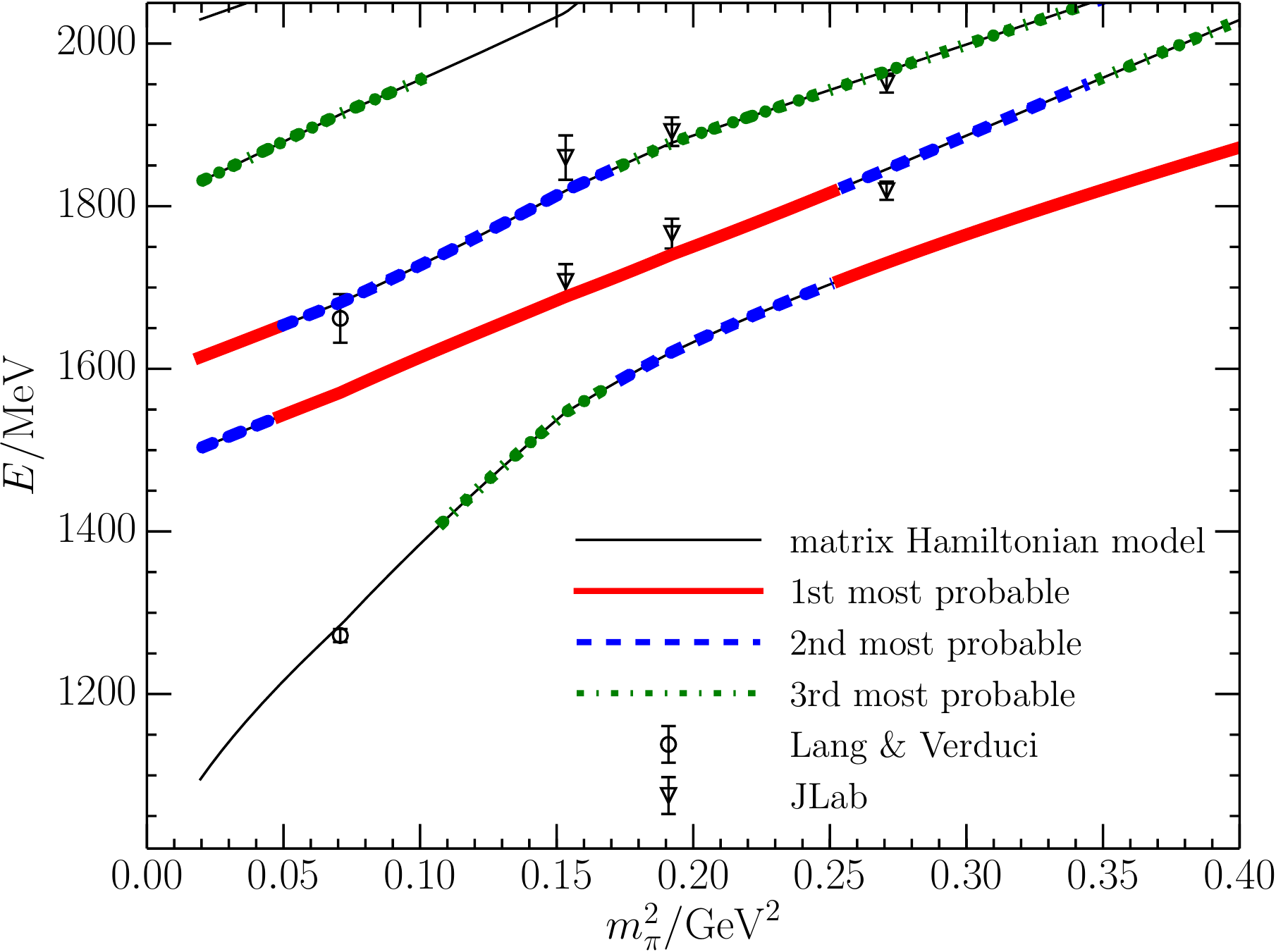}
\includegraphics[width=0.50\textwidth]{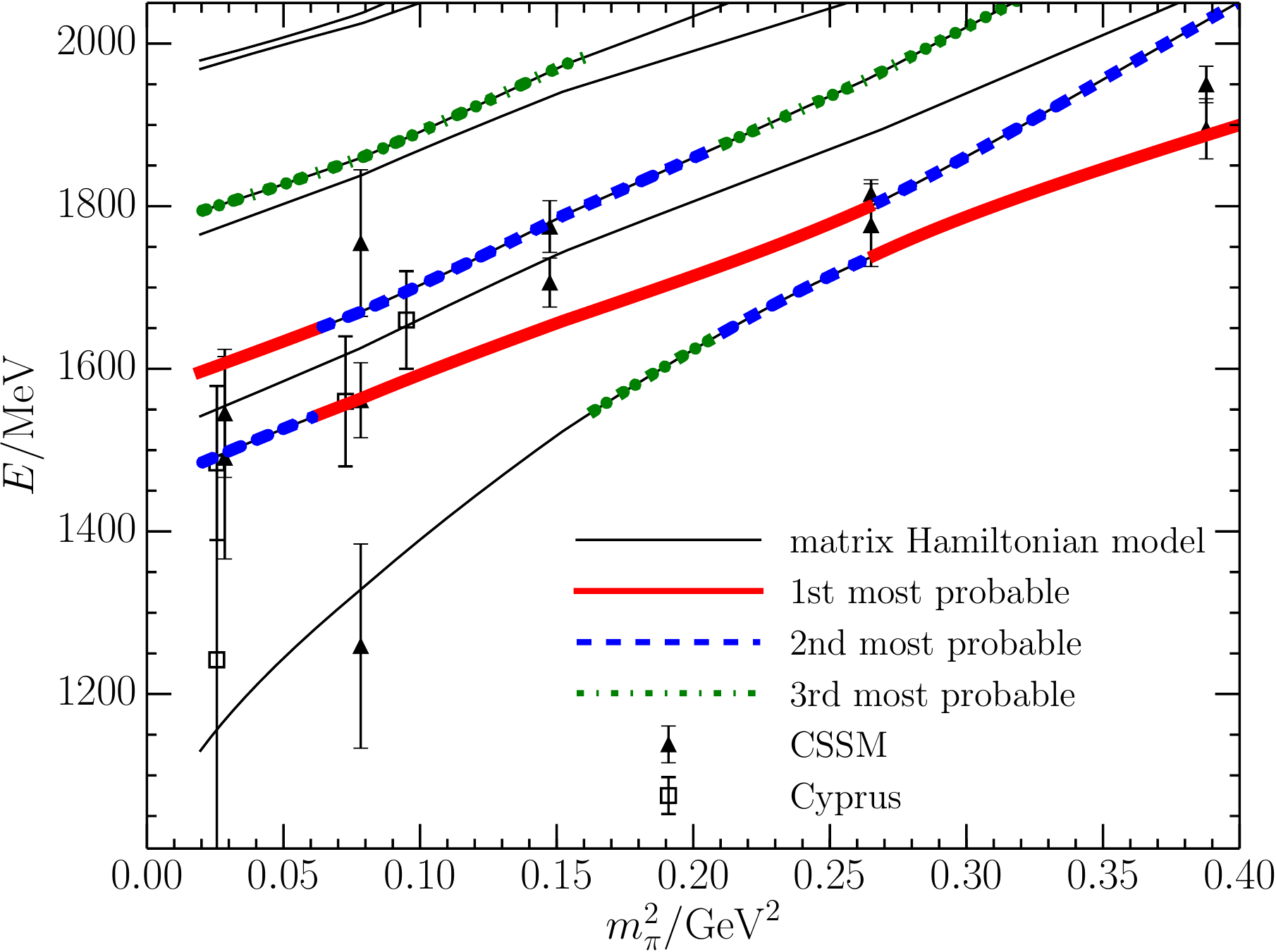}
\caption{\label{fig:Leinweber_HEFT}
 The energy eigenvalues in $J^P=(1/2)^-$ channel of the Hamiltonian for two
 different volumes with $L\simeq 2$fm (left panel) and $L\simeq 3$fm (right panel) at various pion masses,
 taken from Ref.~\cite{Liu:2015ktc}.
 The HEFT results are shown as different line types
 which are compared with the existing lattice data.
 }
\end{figure}
 In the Hamiltonian Effective Field Theory approach, one constructs an
 effective Hamiltonian starting from the non-interacting bare Fock states of relevant hadrons
 and parameterize their interactions using phenomenologically known results.
 For example, one could parameterize the interacting part of the Hamiltonian
 in terms of form factors and compute the low-energy scattering phase shifts
 which are compared with the known experimental results. Demanding that these scattering phase shifts to
 agree with the experiments will constrain the undetermined parameters appearing in
 the effective Hamiltonian. After this procedure, the same effective Hamiltonian
 can be diagonalized numerically in a finite box with a particular chosen volume,
 yielding the eigenvalues $E_\alpha$ which can be compared directly
 with the relevant results from corresponding lattice simulations.

 In Fig.~\ref{fig:Leinweber_HEFT}, we have shown the results from HEFT~\cite{Liu:2015ktc}
 of the $N^*(1535)$ (the negative parity excitation of the nucleon) spectrum
 as a function of the pion mass value squared. Different line types are
 the HEFT results which are compared with relevant lattice data (points with error-bars)
 at two different volumes one with $L\sim 2$fm (left panel) and one with $L\sim 3$fm (right panel).
 Different line types and colors will illustrate the composition of the
 energy eigenstate in terms of the original bare Fock states. Interested readers
 can referred to Ref.~\cite{Liu:2015ktc} for further explanations of their color coding.
 In general, the HEFT can describe the lattice data rather well.

 Some authors of Ref.~\cite{Liu:2015ktc} also believe that, using
 this approach, they could also resolve the Roper resonance~\cite{Leinweber:2015kyz},
 the first excited nucleon state in $J^P=(1/2)^+$ channel.
 There has been a puzzle in this channel for some years, namely the
 Roper turns out to be much higher than its physical value for most of the
 lattice computations except for those using overlap fermions. It is known that
 the overlap fermions, though quite expensive in terms of simulation,
 has the best chiral behavior which is considered to be the major reason for this
 discrepancy~\cite{Liu:2014jua}. The authors of Ref.~\cite{Leinweber:2015kyz} thus
 claim that they have clarified the puzzle using HEFT, since the $E_\alpha$'s that
 are measured on a particular lattice are not exactly the mass values themselves,
 but rather have to be converted to the values quoted in the experiments. However,
 people using overlap fermions in their simulations do not agree,
 see e.g. K.F.~Liu's review on this subject~\cite{Liu:2014jua,Liu:2016rwa}.
 But it is fair to say that we are getting closer to the final solution
 of this puzzle than a few years ago.

 Let us now come to the so-called HAL QCD method.
 The HAL QCD method utilizes the so-called Nambu-Bethe-Salpeter wavefunction that
 can be directly measured on the lattice.
 The HAL QCD collaboration has utilized this method over the years
 in the study of nuclear force and also in the process of baryon-baryon scattering.
 I refer to Savage's review~\cite{Savage:2016} for specific discussions on these issues.
 Recently, HAL QCD also utilized their method to analyze the nature of
 the $Z_c(3900)$ structure~\cite{Ikeda:2016zwx} which will be discussed
 in subsection~\ref{subsec:XYZ}.

 The Optical Potential (OP) method~\cite{Agadjanov:2016mao} is relatively new
 which attempts to measure the optical potential directly on the lattice.
 It is a very appealing approach that could evade some of the complications
 that will be present in L\"uscher approach. Using synthetic data,
 the authors of Ref.~\cite{Agadjanov:2016mao} have shown the successful application of
 this method . However, it remains to be seen how this method is implemented in real lattice simulations.
 On the theoretical side, it is also tempting to further understand the relation
 of this method with the other methods, say the HAL QCD method mentioned above.

\section{Recent progress in lattice spectroscopy}
\label{sec:real_stuff}

 In this section, I will go over recent progress in the field of lattice
 spectroscopy computations. I will start from the more conventional ones and then
 move on to single-channel scattering processes for the light mesons and the charmed
 mesons, the latter is intimately related to the newly discovered $XYZ$ particles.
 Focus will be put on the charmed sector, though other exotic structures
 will also be mentioned. Scattering processes involving baryons will be
 covered by Savage's review~\cite{Savage:2016}.

 \subsection{Proton Neutron mass splitting}
 \label{subsec:pn-mass-splitting}

 Lattice QCD have come to a stage that can address the subtle and difficult problem
 of proton-neutron mass splitting, which requires proper treatment of both QCD and QED on the lattice.
 This mass splitting is tiny, roughly 0.14\% of its average mass value, and has far-reaching
 phenomenological consequences which concerns the very existence and stability of the usual baryonic matter.
 As we will see, it is rather subtly fine-tuned in terms of basic parameters of QCD and QED.
 Two groups, namely Budapest-Marseille-Wuppertal Collaboration (BMW)
 and the QCDSF-UKQCD Collaboration have made attempts towards this goal in the past few years
 which will be recapitulated in the following.

 \begin{figure}[htb]
\includegraphics[height=0.3\textheight,width=0.80\textwidth]{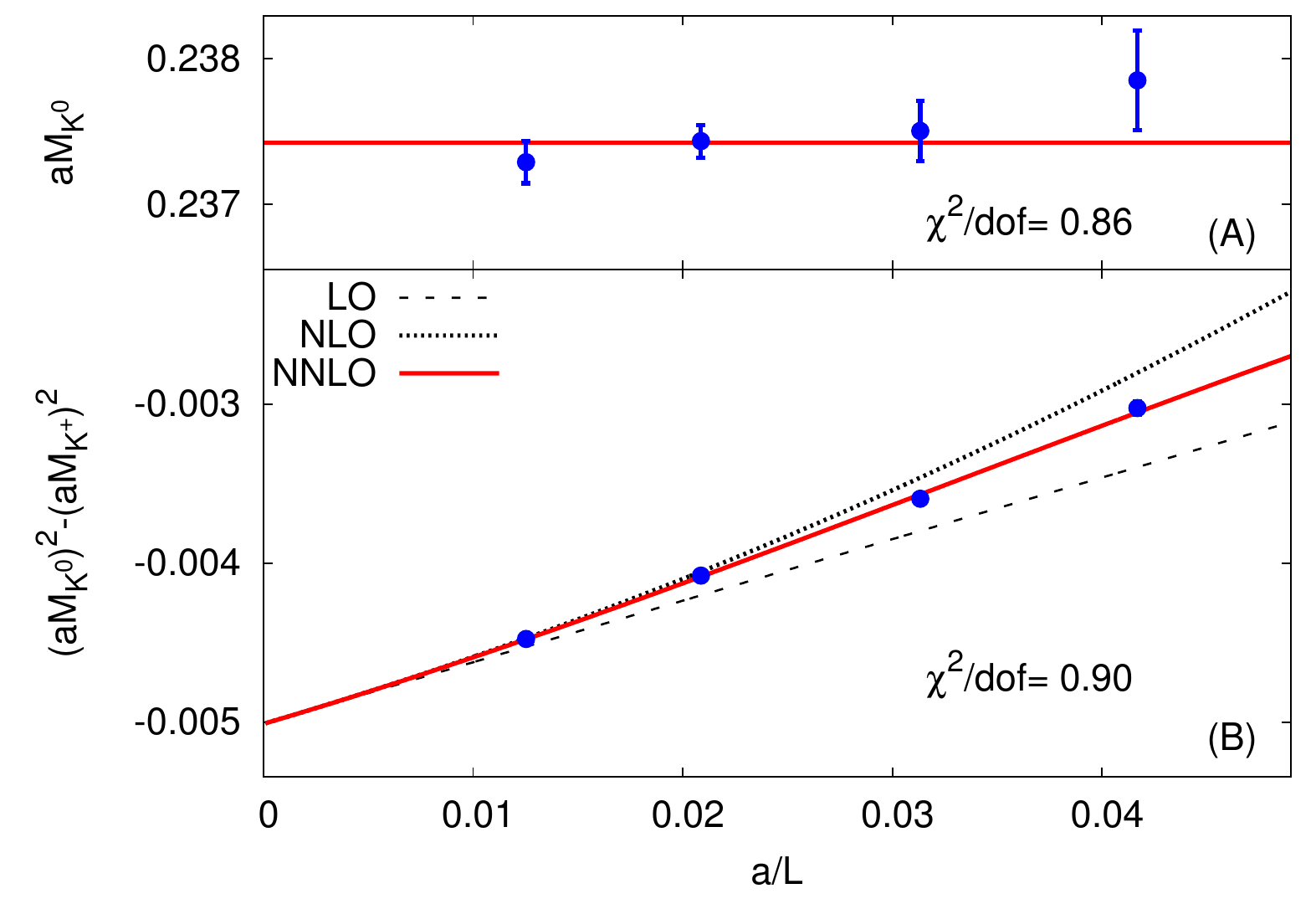}
\caption{\label{fig:BMW_finitevolume}
The finite volume dependence of the neutral and the charged kaon masses from
Ref.~\cite{Borsanyi:2014jba}.}
\end{figure}
 The BMW group simulated $1+1+1+1$ flavor QCD+QED and study the finite volume corrections
 that arise due to the long-range nature of the electromagnetic interaction.
 They compared the the so-called $QED_{L}$~\cite{Hayakawa:2008an,Davoudi:2014qua} and
 the $QED_{TL}$ prescription~\cite{Borsanyi:2014jba}.
 Due to its long range feature, treatment of the electromagnetic field in a finite volume needs
 special care. In the literature, there have been the $QED_{TL}$ and $QED_L$ prescriptions that can be
 applied to the electromagnetic fields.
 The $QED_{TL}$ prescription basically fixed the zero-momentum mode
 of the field $A_\mu(x)$ to zero which violates the reflective positivity, since
 this constraint on $\tilde{A}_\mu(0)$ involves time slices that are far apart in real space.
 It also introduces potential dangers since the charged particle propagators are ill-defined.
 The $QED_{L}$ prescription only do so for the spatial-momentum zero mode but for every
 individual time-slice. Therefore it does not involve far separated time slices and thereby
 preserve reflective positivity. It however violates cubic symmetry which can be viewed
 as a finite volume effect. The BMW group compares  the shift in the pole mass
 of a point particle in both $QED$ perturbation theory and
 numerical simulations and show that the $QED_{L}$ prescription looks normal.
 Fig.~\ref{fig:BMW_finitevolume} illustrates the finite volume dependence of the
 neutral and the charged kaon masses. It is seen that the neutral kaon shows little (exponentially small)
 volume dependence while the charge one shows considerable finite-volume corrections
 that are well described by theoretical expectations.

 \begin{figure}[tbh]
\includegraphics[width=0.50\textwidth]{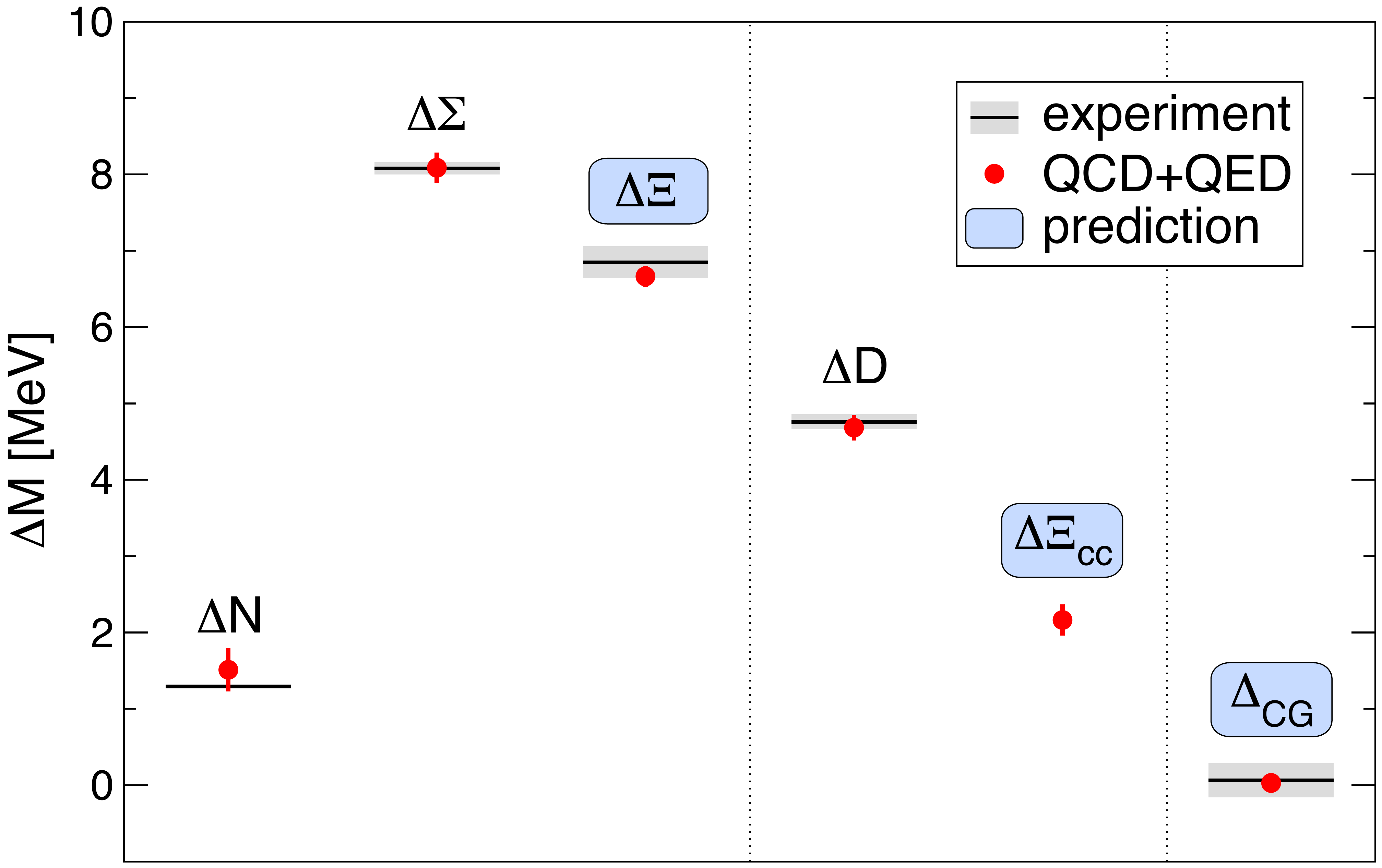}
\includegraphics[width=0.50\textwidth]{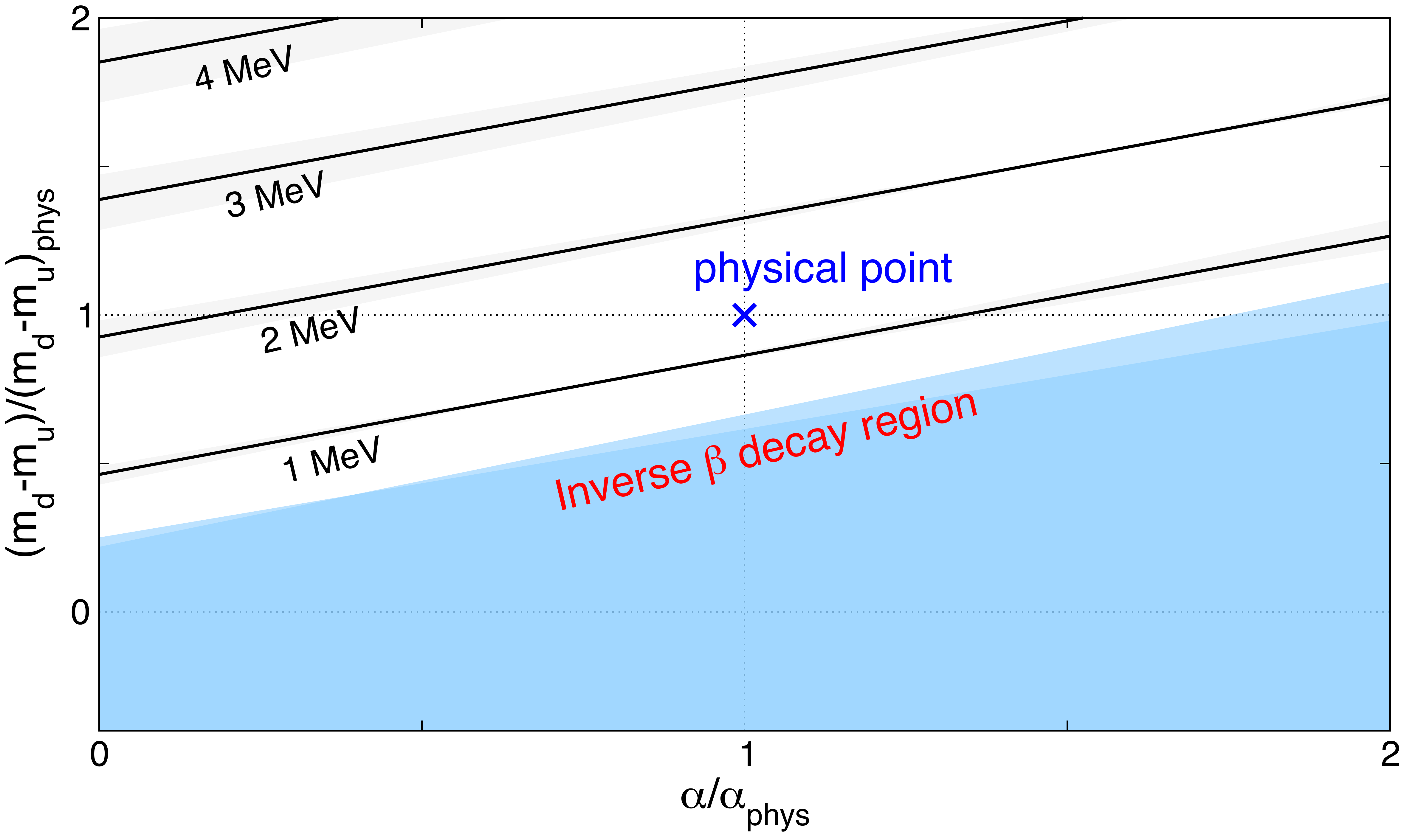}
\caption{\label{fig:BMW_masssplittings}
The mass splittings of various hadrons (left panel) and the
corresponding allowed regions (right panel) in terms of quark masses and fine structure constant~\cite{Borsanyi:2014jba}.}
\end{figure}
 After treating the finite volume corrections carefully,
 BMW collaboration proceeds to obtain the mass splittings for various hadrons:
 the nucleon $\Delta N$, the $\Sigma$ baryon,
 the $\Xi$ baryon, the $D$ meson, the doubly charmed $\Xi$ baryon
 and also the so-called Coleman-Glashow mass difference $\Delta_{CG}$, which is
 supposed to vanish if the Coleman-Glashow relation is valid~\cite{Coleman:1961jn}.
 These results are shown in the left panel of
 Fig.~\ref{fig:BMW_masssplittings}. It is seen that the agreement with the
 existing experimental data is excellent. For example, the tiny mass splitting
 between the proton and the neutron is obtained. Their results even predict
 other mass splittings that have not yet been measured experimentally.

 It is useful to separate the QCD and QED contributions to various
 mass splittings: $\Delta M_X=\Delta_{QCD}M_X+\Delta_{QED}M_X$,
 where $\Delta_{QCD}M_X$ is proportional to $\delta m=m_d-m_u$ while
 $\Delta_{QED}M_X$ is proportional to $\alpha_{EM}$. This separation
 of course is ambiguous at the order of $\calO(\alpha\delta m)$.
 BMW argued that, to a good approximation, the mass splitting of the $\Sigma$ baryon comes
 solely from QCD. After fitting their data sets, the mass splitting between
 the neutron and proton can be separated into QCD and QED contributions.
 This is illustrated in the right panel of Fig.~\ref{fig:BMW_masssplittings}.
 The horizontal and vertical axis are essentially $\alpha\equiv\alpha_{EM}$ (the fine-structure
 constant due to electromagnetism)
 and $\delta m=m_d-m_u$ (the difference of the down and up quark masses) measured
 by their corresponding physical values. The color shaded region is ruled out due to the inverse $\beta$ decay.
 The contour lines indicate constant $m_N-m_p$ values and the true physical
 point is indicated by a cross. This figure illustrates the subtle balance
 between the QCD and the QED contributions to neutron-proton mass splitting.

\begin{figure}[tbh]
  \hspace{-8mm}
\mbox{\begin{minipage}[b]{200pt}
\includegraphics[width=1.30\textwidth]{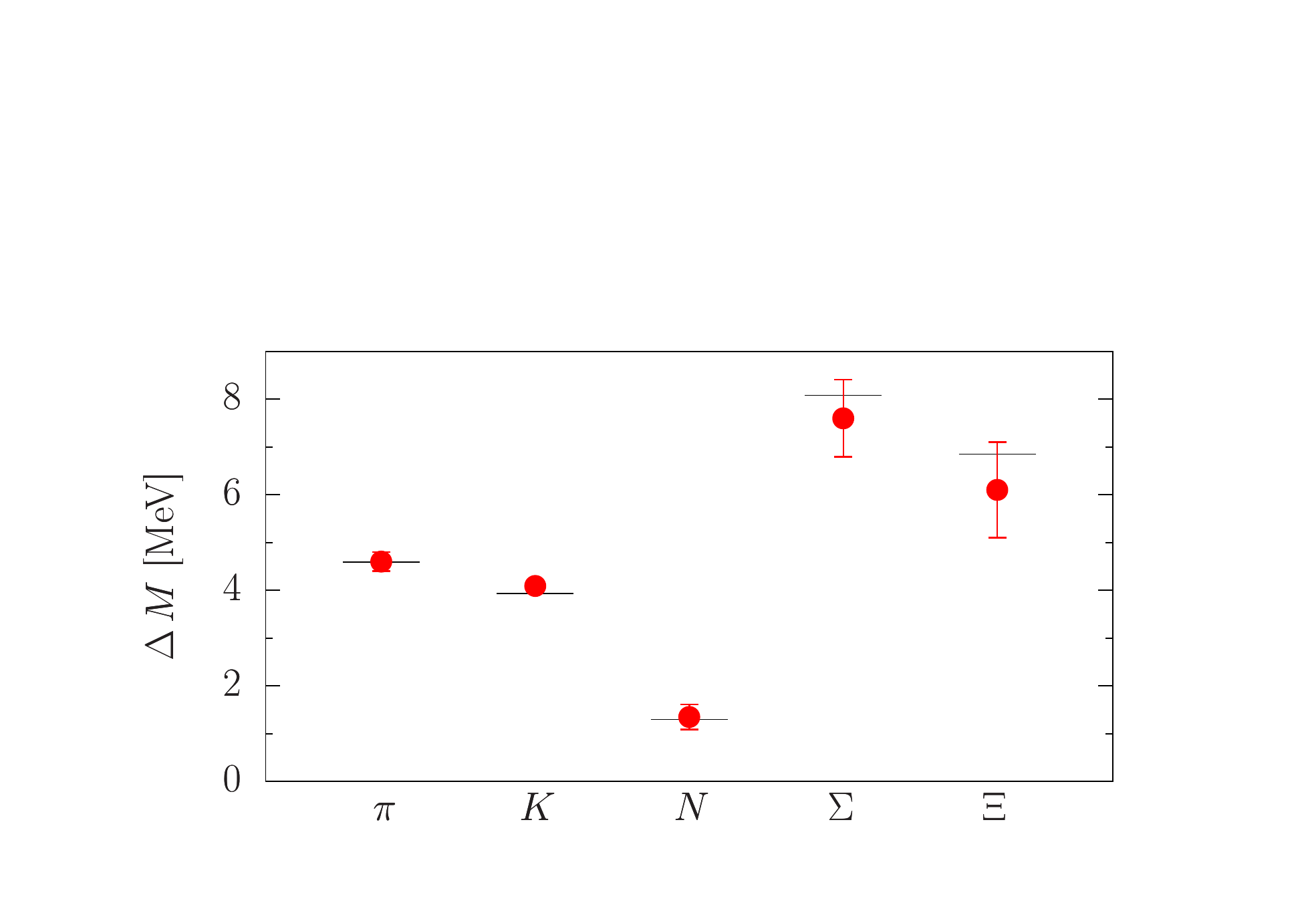}
\end{minipage}}
\hspace{-2mm}
\raisebox{-5mm}{\mbox{\begin{minipage}[b]{200pt}
\includegraphics[width=1.30\textwidth]{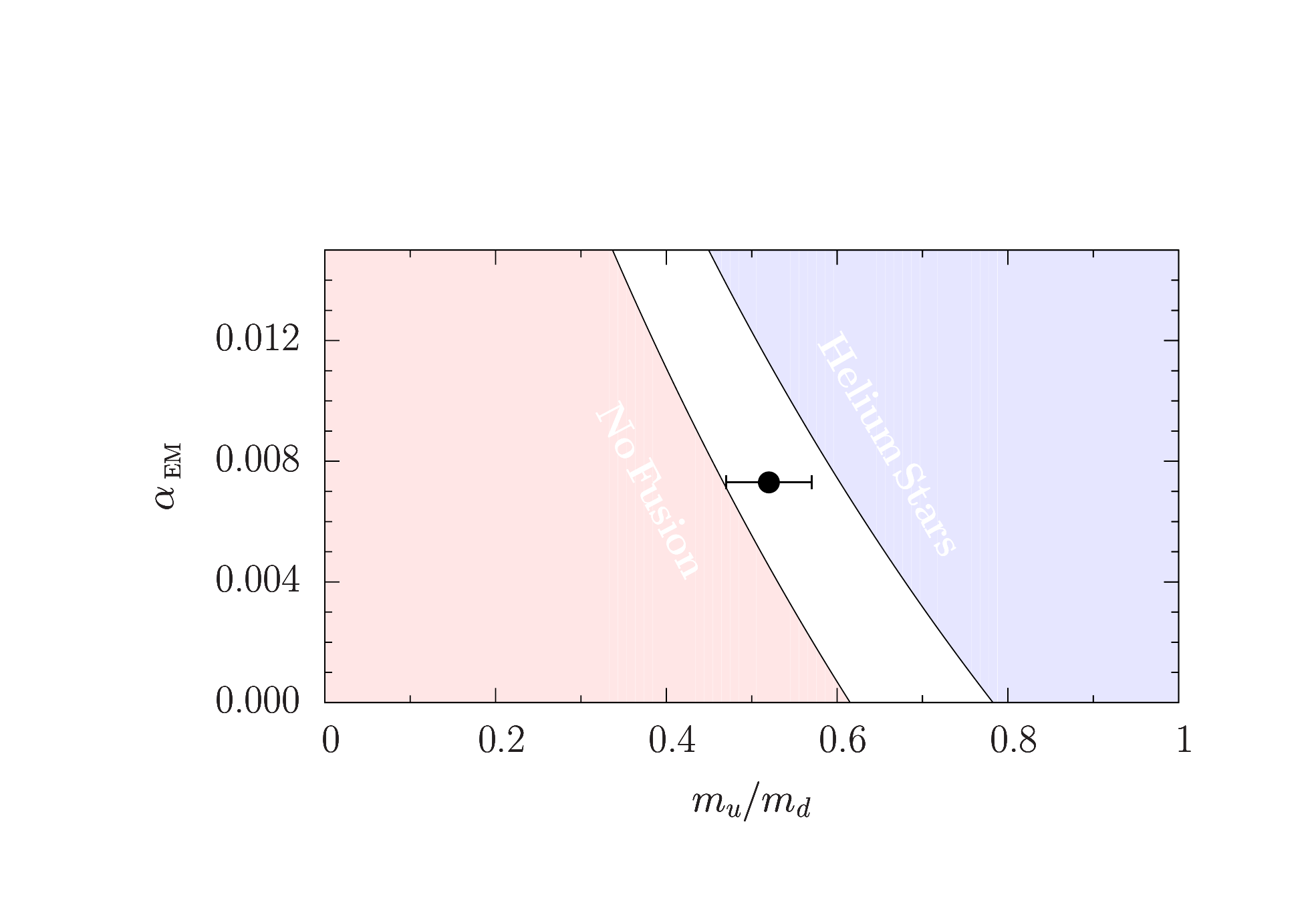}
\end{minipage}}}
\caption{\label{fig:QCDSF_masssplittings}
The final mass splittings for various hadrons (left panel) and
the QCD (in terms of $m_u/m_d$) and QED contributions to the
neutron-proton mass splittings~\cite{Horsley:2015vla}.}
\end{figure}

 The QCDSF-UKQCD collaboration made a similar attempt~\cite{Horsley:2013qka,Horsley:2015eaa,Horsley:2015vla}.
 They choose to tackle the problem from the so-called $SU(3)$ symmetric point of pure QCD simulations
 where all three quarks have the same masses.
 The so-called Dashen scheme was adopted and the conversion to other schemes was also discussed.
 They studied the octet baryons, the octet mesons, quark masses as well as
 vacuum structures. Using Dashen scheme, QCDSF is able to separate the QCD and QED
 contributions to the neutron-proton mass difference. They have also checked
 the Coleman-Glashow relation~\cite{Coleman:1961jn}.
 Fig.~\ref{fig:QCDSF_masssplittings} shows the summary of their results.
 In the left panel, the mass splittings for the pion ($\pi$), kaon ($K$),
 nucleon ($N$), $\Sigma$ baryon and $\Xi$ baryon are compared with the experimental
 values which are indicated by the horizontal bars. The overall agreement
 is impressive. In the right panel, the equivalent of right panel of
 Fig.~\ref{fig:BMW_masssplittings} is shown but plotted in a different way.
 The neutron-proton mass difference is
 plotted by separating the contribution from QCD (indicated by the parameter $m_u/m_d$)
 and QED (indicated by the parameter $\alpha_{EM}$). The two color shaded regions
 are actually forbidden from cosmological point of view since there is either no fusion
 or no regular star formation.

 \subsection{Simulation at the physical point}
 \label{subsec:physical-point}

  \begin{figure}[tbh]
\includegraphics[width=0.49\textwidth]{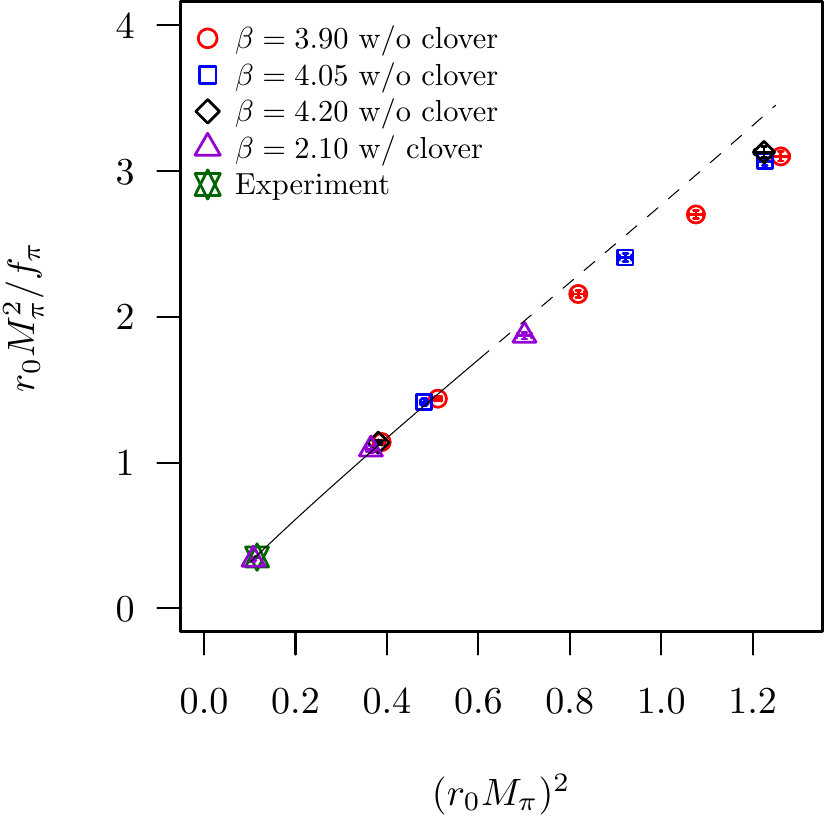}
\includegraphics[width=0.49\textwidth]{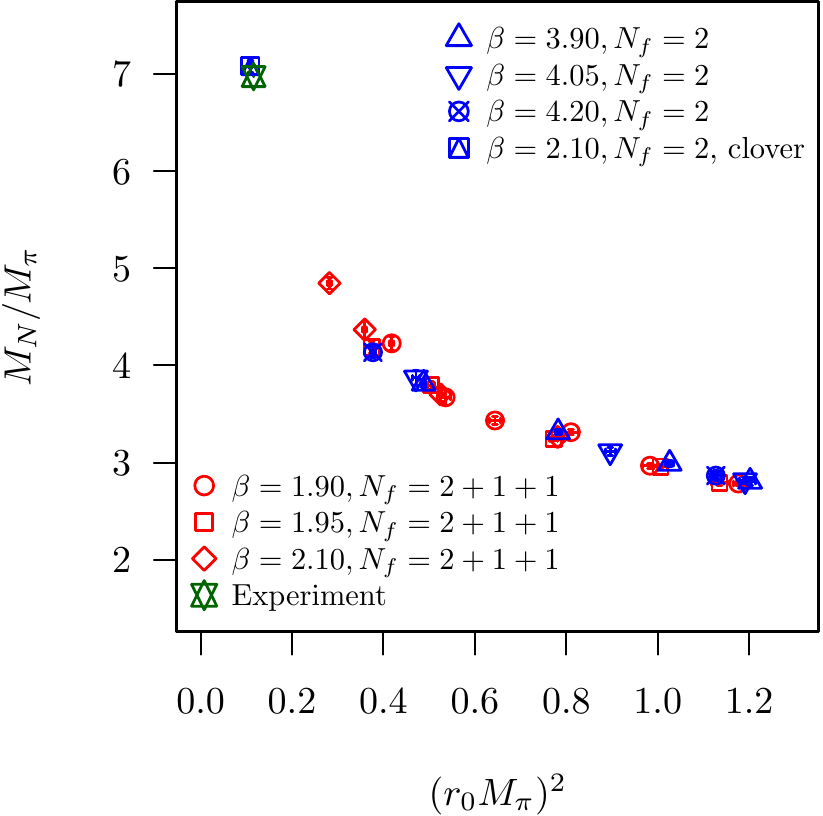}
\caption{\label{fig:ETMC_physical}
Typical chiral and continuum extrapolation in physical pion mass simulation
of ETMC~\cite{Abdel-Rehim:2015pwa}. Left panel shows the quantity $r_0m^2_\pi/f_\pi$ while the right
panel shows the mass ratio of the nucleon to the pion. The physical simulation point
corresponds to the open purple triangle in the left and the blue square in
the right panel, both of which having the smallest value of $(r_0m_\pi)^2$. }
\end{figure}
 Apart from the above mentioned lattice computations, the ETM collaboration performed
 a simulation at the physical pion mass~\cite{Abdel-Rehim:2015pwa}. The simulation was
 done at $\beta=2.10$ with  a clover term added. Various quantities
 have been computed and compared with the experimental results and other lattice results.
 These includes, meson and baryon masses, decay constants and their ratios, quark masses, etc.
 Chiral behavior is inspected with care.
 Fig.~\ref{fig:ETMC_physical} shows a typical chiral behavior of $r_0m^2_\pi/f_\pi$ (left panel)
 and $m_N/m_\pi$ (right panel) vs. $(r_0m_\pi)^2$ where $r_0$ being the Sommer scale.
 It is seen that the direct simulated results, the purple triangle in the left panel and
 the blue square in the right, agree very well with the
 expected result obtained from chiral extrapolation (line in the left panel) and the
 corresponding experimental results (the green hexagrams in both panels).
 In the right panel, some of the $2+1+1$ results from ETMC (the red points) are also shown for reference.

 In Ref.~\cite{Abdel-Rehim:2015pwa}, ETMC also studied other important hadronic quantities:
 various hadron masses and decay constants (and their ratios), quark masses and
 the hadronic contributions to the lepton anomalous magnetic moments.
 They have compared their lattice results with the experimental ones
 and good agreements are found.

%  \begin{figure}[tbh]
%\includegraphics[height=0.29\textheight]{ETMC_summary.pdf}
%\caption{\label{fig:ETMC_summary}
%Summary of the lattice results.}
%\end{figure}

\subsection{Single-channel scattering of light mesons}
\label{subsec:pipi-scattering}

 As mentioned in the previous section, lattice computations utilizing
 L\"uscher's formalism have matured in recent years and there have
 been a number of such calculations in the past year. I will go through the
 light meson scattering computations first.

  \begin{figure}[tbh]
\includegraphics[width=0.89\textwidth]{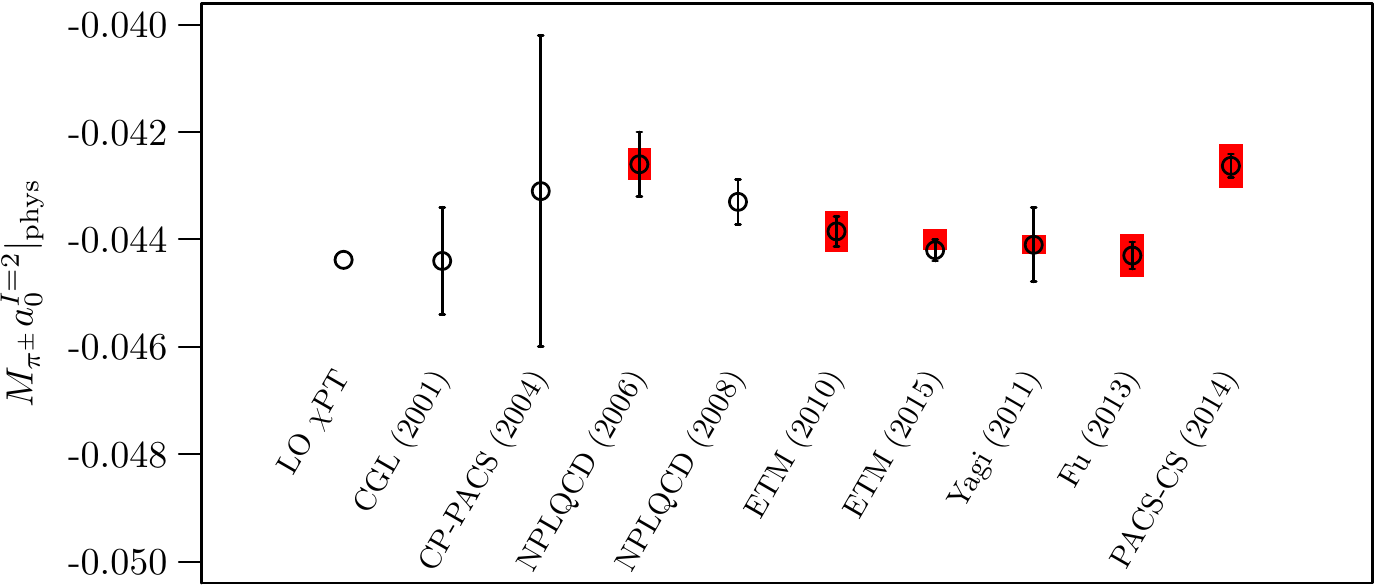}
\caption{\label{fig:ETMC_pipia0}
Comparison of ETMC~\cite{Helmes:2015gla}'s result, marked as ETM (2015), for $a^{I=2}_0$ with existing results in the literature. The two points to the far
left correspond to leading order chiral perturbation w/o dispersion analysis from the continuum.
The rest are all lattice results. Red bands indicate the systematic errors where available. }
\end{figure}
 The ETM Collaboration has been studying the low-energy pion-pion
 scattering processes using their $2+1+1$ flavor ensembles~\cite{Helmes:2015gla,Helmes:2015uhe}.
 In Ref.~\cite{Helmes:2015gla}, the $\pi\pi$ scattering length in $I=2$ channel,
 denoted by $a^{I=2}_0$, is obtained in the continuum and physical pion mass limit.
 As we all know now, scattering length is a very important low-energy quantity
 that will enter many effective field theory analysis. For example, the $\pi N$ scattering length
 might be the crucial point to resolve the discrepancy for the sigma term
 between the existing lattice computations and the dispersion relations~\cite{Hoferichter:2016ocj}.
 Therefore, pion-pion scattering length serves as a good benchmark quantity for lattice
 computations.

 The result of $M_{\pi^\pm}a^{I=2}_0$ from Ref.~\cite{Helmes:2015gla}
 is plotted in Fig.~\ref{fig:ETMC_pipia0} (marked as ETM (2015)) together
 with those from chiral perturbation theory and other previous lattice computations.
 The two points to the left are from continuum determinations,
 either using leading order chiral perturbation theory (marked as LO $\chi PT$) or together
 with dispersion relations (marked as CGL (2001)). All the rest are
 from various lattice computations performed with the chiral and
 continuum extrapolations. The error-bars stand for the
 statistical errors while the red shaded bands on the points indicate the systematic
 errors where available. Careful analysis of the statistical and systematic errors have been
 performed in the study of Ref.~\cite{Helmes:2015gla} and similar studies in other channels of
 pion-pion and pion-kaon scattering are under way.

  \begin{figure}[tbh]
\includegraphics[width=0.43\textwidth]{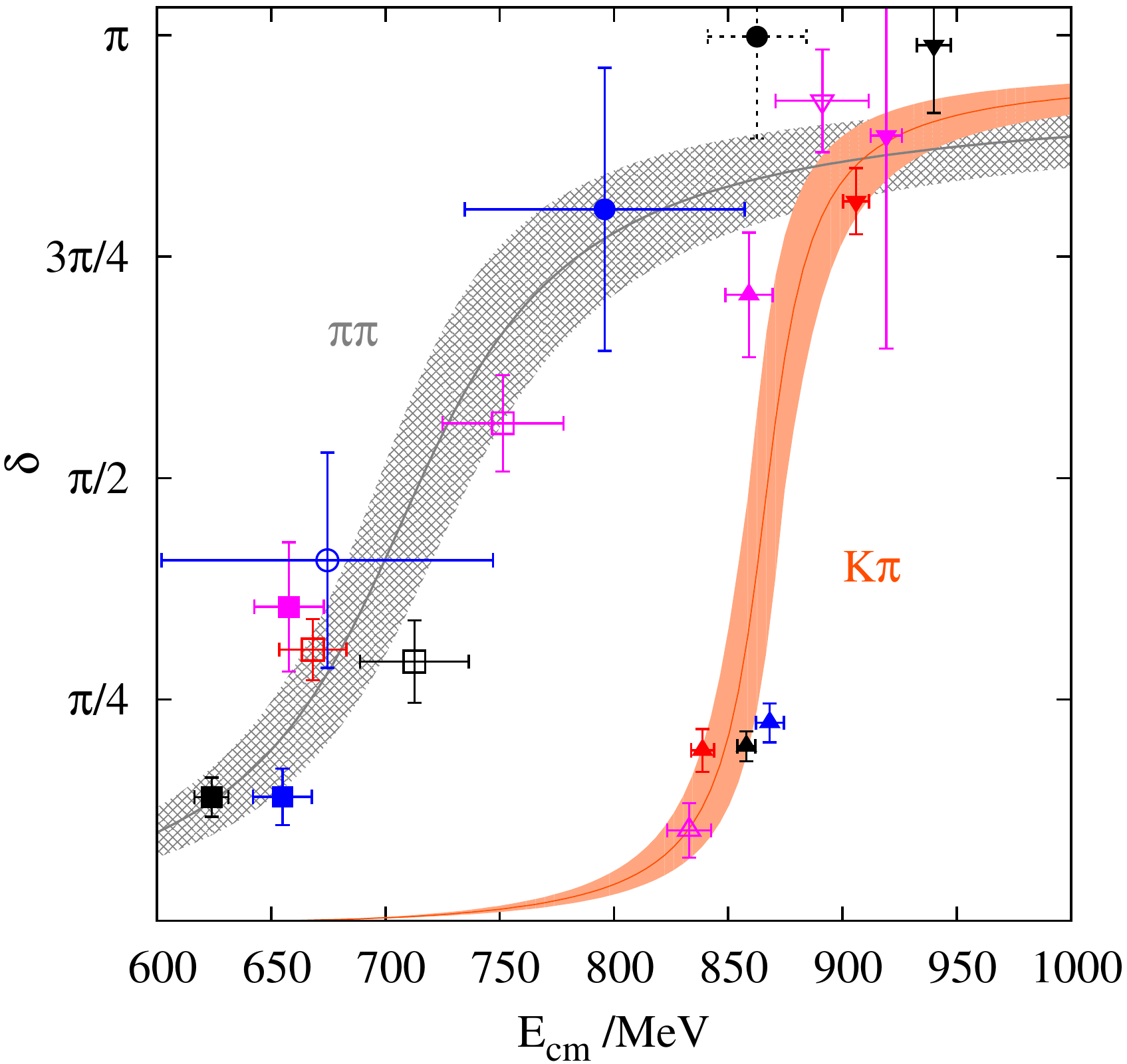}
\includegraphics[width=0.56\textwidth]{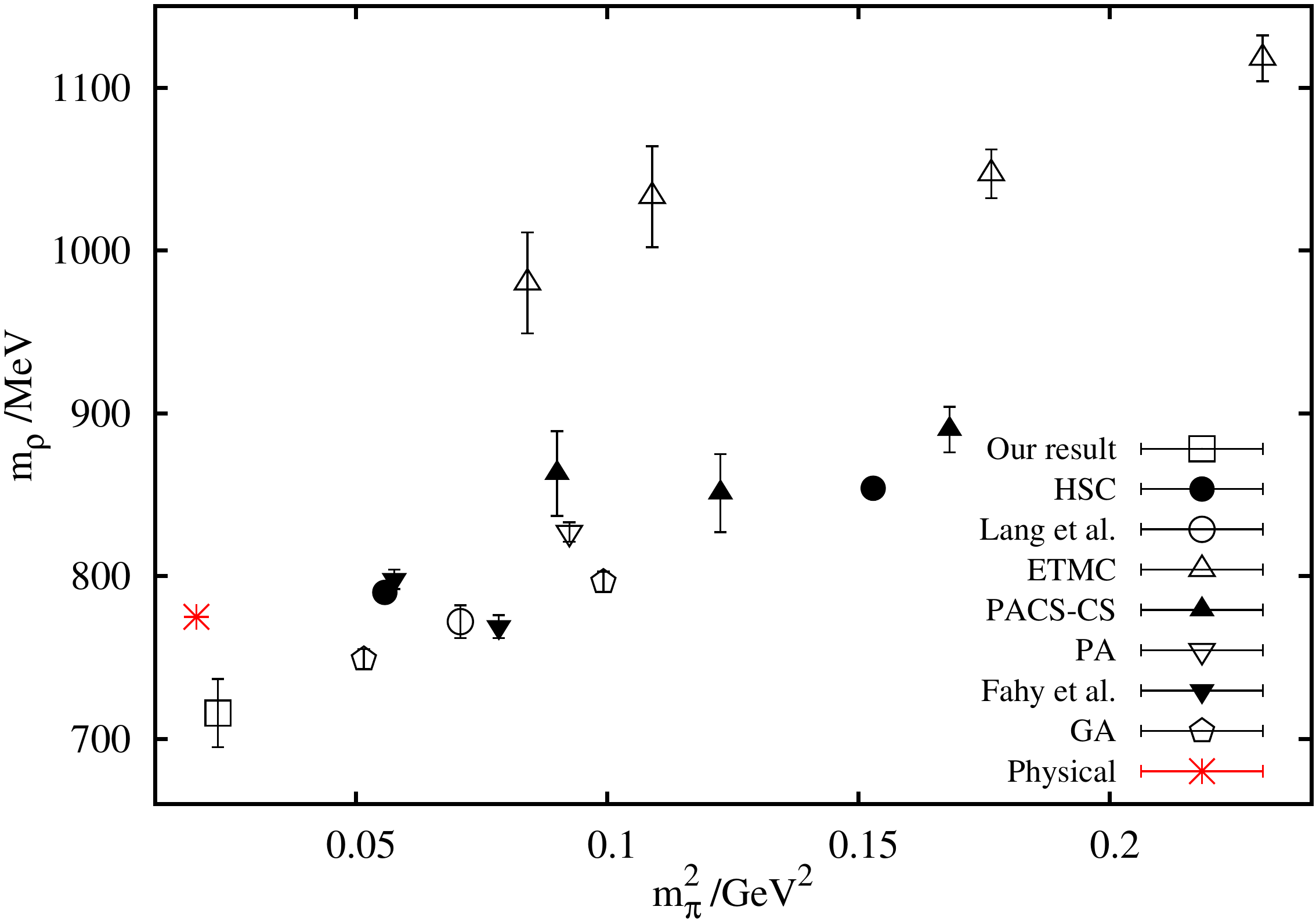}
\caption{\label{fig:RQCD_rho}
 Left panel shows the phase shifts of $\pi\pi$ and $K\pi$ in the corresponding channel form RQCD computation~\cite{Bali:2015gji}. The right panel illustrates the summary of various lattice
 determinations for $m_\rho$ as a function of pion mass.}
\end{figure}
 In recent years, the computation of the rho resonance becomes not only feasible but
 also quite fashionable, illustrating our understanding of the resonance nature of
 $\rho$ meson. In the past year or so, there have been a number of computations of this using different
 techniques and with various fermion realizations~\cite{Bali:2015gji,Bulava:2016mks,Wilson:2015dqa,Guo:2016zos}.
 Some of these computations have already been quite systematic in the sense that
 a series of lattice ensembles
 have been utilized in the computation,
 enabling one to estimate various systematics in a reliable fashion.
 As an example, in Fig.~\ref{fig:RQCD_rho} we show the phase shifts obtained by the RQCD Collaboration
 using almost physical pion mass ($m_\pi\simeq 150MeV$) with $N_f=2$ Wilson fermions.
 They studied both $\pi\pi$ and $K\pi$ scattering in the $p$-wave and the scattering phase shifts are shown
 in the left panel of  Fig.~\ref{fig:RQCD_rho}, depicting the resonance structure
 of the rho (the $\pi\pi$ channel) and the $K^*$ (the $K\pi$ channel), respectively.
 In the right panel, they summarize
 the Breit-Wigner mass of the $\rho$ from various lattice groups.
 It is seen that most $N_f=2$ results
 (the open symbols) undershoot the corresponding physical value (the red star) and
 this is likely due to the quenching of the strange quark,
 as the authors of Ref.~\cite{Guo:2016zos} have argued.
 The filled symbols are from $N_f=2+1$ simulations and they seem
 to converge to the physical result rather well.

 To summarize, concerning the single-channel scattering of light mesons,
 physical properties have been studied with good accuracy and control.

\subsection{Single-channel scattering of charmed mesons and the XYZ particles}
\label{subsec:XYZ}

 In recent years, a handful of near-threshold structures have been observed in
 the experiments. This happened in both the bottom and the charm sector.
 These structures, though the nature of them remains to be clarified, have been called the $XYZ$ particles.
 A wealth of phenomenological explanations have been put forward
 including: conventional quarkonium, molecular states, tetra-quark states, etc.,
 see e.g. Ref.~\cite{Chen:2016qju} and references therein.
 Lattice studies can also shed some light on these possible explanations.
 Below, I will focus on lattice studies on some of the $Z_c$ states.

 One of these $Z_c$ state is $Z_c(3900)$, observed by BESIII,
 Belle and CLEO-c collaborations~\cite{Ablikim:2013mio,Liu:2013dau,Xiao:2013iha}
 whose quantum numbers are: $I^G(J^{PC})=1^+(1^{+-})$.
 The state is just around the $\bar{D}D^*$ threshold and interacts strongly
 with the $\bar{D}D^*$ final states. Therefore, it
 was naturally conjectured to be a loosely bound state of the two relevant charmed mesons.
 However, things might be more complicated than this.
 There are in fact quite a number of other thresholds below that
 of $\bar{D}D^*$, e.g. $\eta_c\rho$, $\pi J/\psi$ etc., therefore
 multi-channel effects might be relevant here. This is particularly complicated
 for lattice computations since it relies on the operator building process
 as we described in section~\ref{sec:theory}. More importantly, if one would like to pursue
 the usual L\"uscher's approach, one in principle has to deal with
 a multi-channel situation, much more complicated than the single-channel
 version as in the $\pi\pi$ sector which we described in subsection~\ref{subsec:pipi-scattering}.
 This has been realized for quite some time,  see e.g. Ref.~\cite{Prelovsek:2014zga}
 and references therein.

  \begin{figure}[tbh]
  \vspace{-40mm}
\resizebox{0.86\textwidth}{!}{\includegraphics{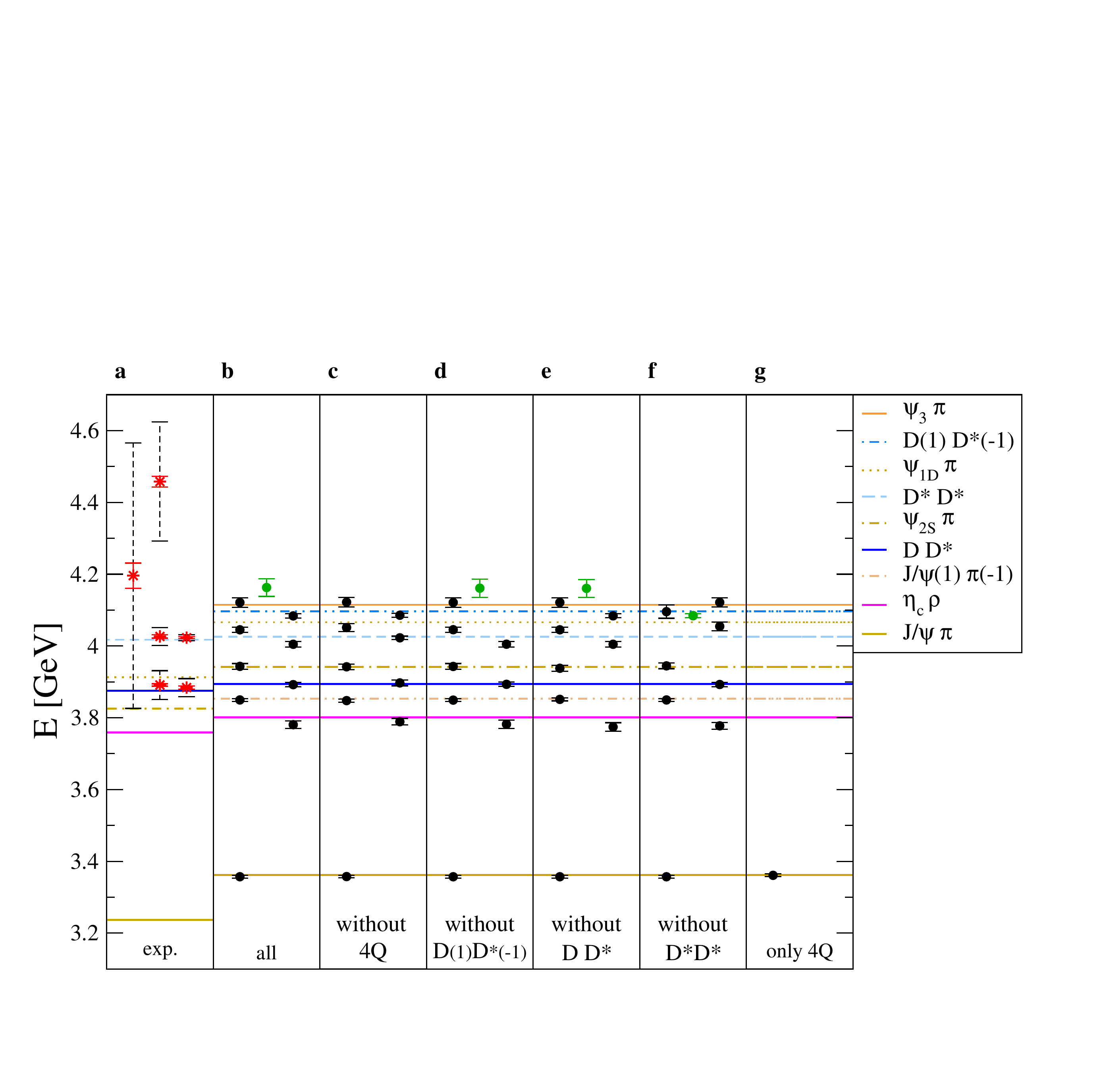}}
\vspace{-10mm}
\caption{\label{fig:sasa_Zc}
 The spectra obtained by Prelovsek {\em et al}, taken from Ref.~\cite{Prelovsek:2014swa}.}
\end{figure}
 Prelovsek {\em et al} studied this problem using one set of $16^3\times 32$
 $N_f=2$ Wilson fermion gauge field ensemble
 that corresponds to $m_\pi\sim 266$MeV. The lattice spacing ($a\sim 0.124$fm) and the
 volume ($L\sim 2$fm) are all fixed.
 However, they used a rather elaborate multi-channel operator basis with distillation,
 a novel smearing technique~\cite{Peardon:2009gh,Morningstar:2011ka}.
 Their operator basis includes the so-called tetra-quark operators as well.
 \footnote{One should keep in mind that these tetra-quark operators
 are in fact related to the two meson operators via Fierz rearrangement.
 Therefore, the two sets of operators are in fact not linearly independent.}
 They obtained the finite volume spectra, namely those $E_\alpha$'s,
 and compare the spectrum with the free two-meson spectra
 and see if an extra state should emerge~\cite{Prelovsek:2014swa}, a strategy
 that had been successfully utilized in their previous search for $X(3872)$~\cite{Prelovsek:2013cra}.
 The summary plot of their spectra is shown in Fig.~\ref{fig:sasa_Zc}
 in which no new exotic state could be identified.
 Therefore, their conclusion is negative, namely no new exotic state is found
 below $4.2$GeV from their lattice computation.

% There have been two major groups working on these, the CLQCD and Prelovsek {\em et al}.

 To simplify the multi-channel nature of the problem,
 China Lattice QCD (CLQCD) attempts to single out the most important channel of the problem and
 proceeds with the single-channel L\"uscher approach.
 Normally, from phenomenological arguments and experimental facts, one knows
 that the newly discovered structure strongly couples to one particular channel.
 For example, for the case of $Z_c(3900)$, the most important channel is $\bar{D}$ and a $D^*$
 while for $Z_c(4025)$ the major channel is $D^*\bar{D}^*$.
 Then, building two-meson operators in these corresponding channels
 with the right quantum numbers will allow us to explore the problem
 within this single-channel approximation.
 CLQCD had estimated the effects of other operators/channels, making sure that they
 do not ruin the major channel correlators~\cite{Chen:2014afa,Chen:2015jwa}.
 Partially twisted boundary conditions have been utilized to fully
 explore the near threshold region.
 Then, using single-channel L\"uscher formalism combined with effective range expansion,
 \be
 \label{eq:ere}
 k\cot\delta(k)={1\over a_0}+{1\over 2}r_0k^2+\cdots\;,
 \ee
 they were able to determine the elastic scattering length $a_0$ and the effective range
 $r_0$ of the two charmed mesons in the near-threshold energy region,
 where $k$ stands for the magnitude of the scattering momentum of the two mesons
 in the center of mass frame.

  \begin{figure}[tbh]
\includegraphics[width=0.49\textwidth]{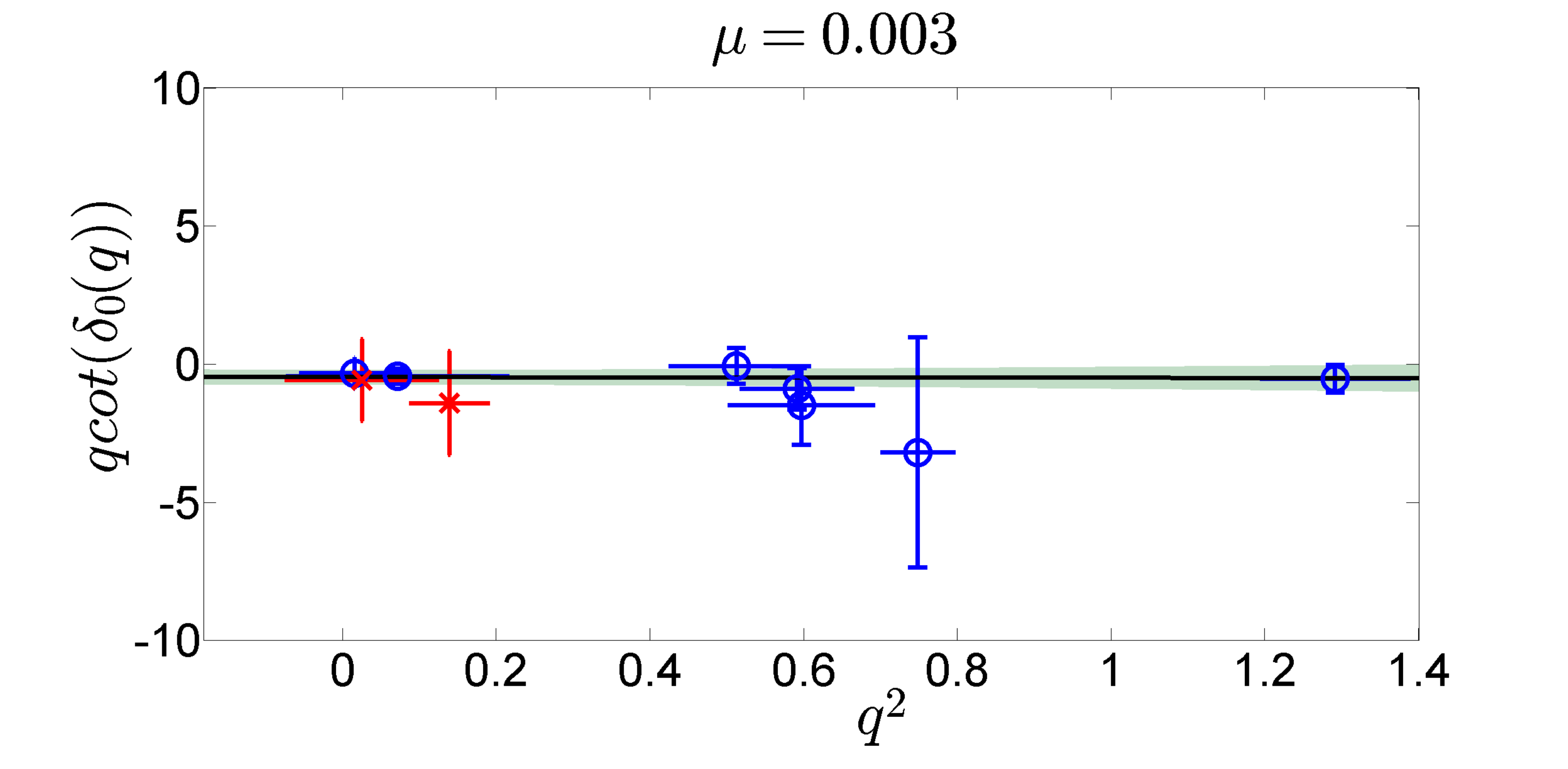}
\includegraphics[width=0.49\textwidth]{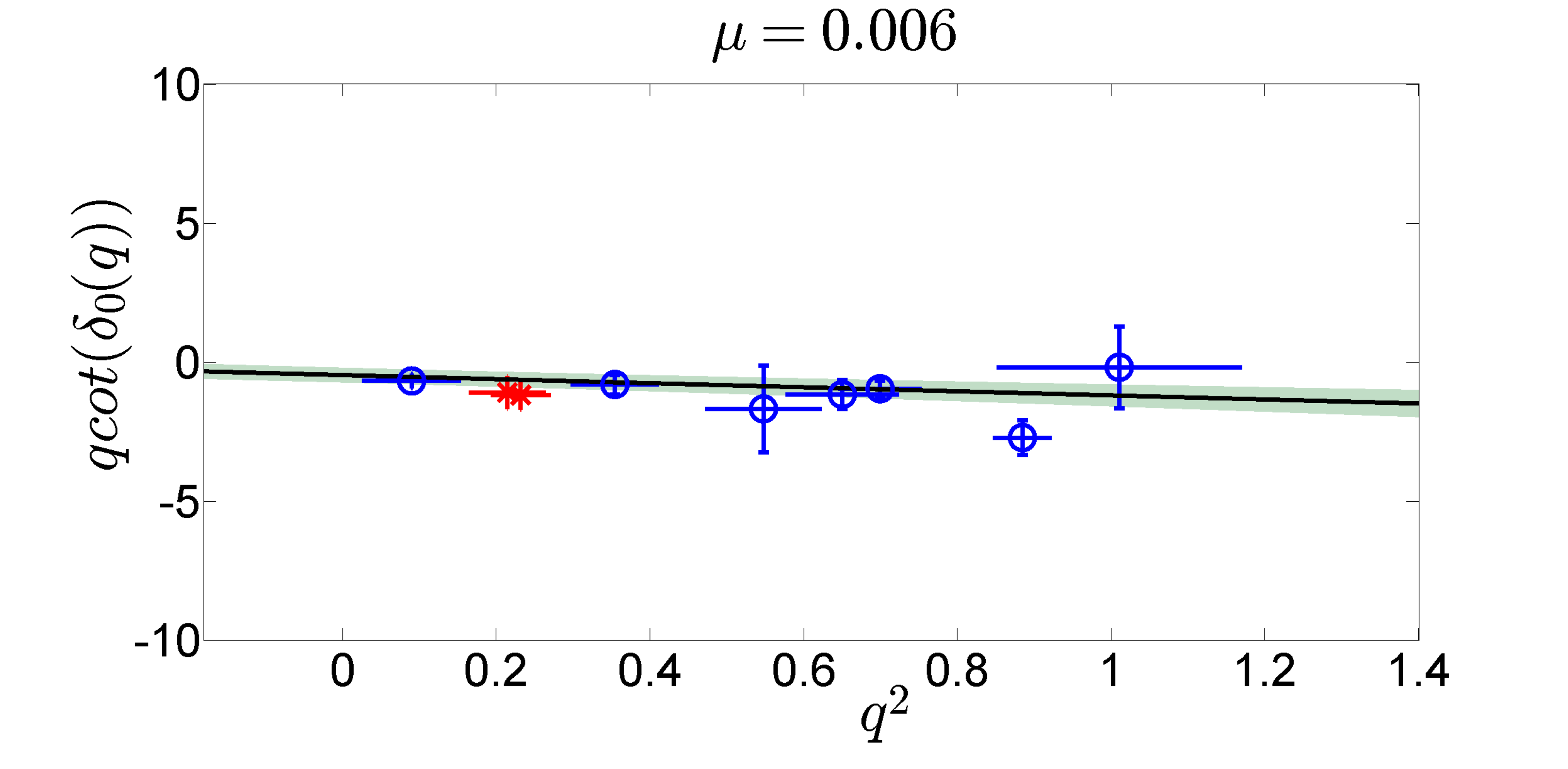}
\caption{\label{fig:CLQCD_Zc}
 The quantity $q\cot(\delta_0(q^2))$ is plotted vs. the dimensionless
 scattering momentum squared, $q^2$, at two different pion mass values,
 taken from Ref.~\cite{Chen:2015jwa}.}
\end{figure}
 CLQCD utilized $N_f=2$ twisted mass configurations with three different pion masses ranging
 from $300$MeV to $485$MeV.
 However, all configurations are at a fixed lattice spacing ($a\sim 0.067$fm) and
 a fixed physical volume ($L\sim 2.1$fm).
 In Fig.~\ref{fig:CLQCD_Zc} the effective range fitting
 from CLQCD's study on $Z_c(4025)$ is illustrated~\cite{Chen:2015jwa}.
 The horizontal axis is proportional to the scattering momentum squared, $q^2=(kL/(2\pi))^2$ while
 the vertical axis shows the quantity $q^2\cot\delta(q^2)$.
 The straight lines are linear fits according to Eq.~\ref{eq:ere}
 near the threshold $q^2=0$ and the intercepts of the straight lines
 basically yield $1/a_0$. Slightly negative scattering lengths have
 been obtained. This is quite analogous to the case of $\pi\pi$ scattering
 in the $I=2$ channel. Their results thus indicate that
 the two charmed mesons seem to have weak repulsive interactions
 and no bound states are found for all three pion mass values in their lattice
 computation. Similar situation has been witnessed in the study of $Z_c(3900)$~\cite{Chen:2014afa}.
 However, this weak repulsion scenario is not universal for all cases of charmed mesons.
 In a similar lattice study on $Z(4430)$, a structure close
 to the threshold of a $D_1$ and $\bar{D}^*$ which was first observed by Belle~\cite{Choi:2007wga}
 back in 2008 and later on verified by LHCb~\cite{Aaij:2014jqa}, CLQCD did find attractive interaction
 between the two charmed mesons in both quenched lattice QCD and in the unquenched case~\cite{Chen:2016lkl}.

 The two different approaches mentioned above are in fact quite complimentary to each other.
 The main conclusion they reach is also similar: No indication from the lattice computation
 has been observed for the state $Z_c(3900)$ and $Z_c(4025)$.
 A similar preliminary study using HISQ lattices by the Fermilab and MILC collaboration also fails
 to identify any indications for these exotic states~\cite{Lee:2014bea}.

 On the other hand, HAL QCD collaboration tackled the problem using their
 HAL QCD approach. They used improved Wilson gauge field configurations PAC-CS
 with $2+1$ dynamical flavors at one lattice spacing ($a\sim 0.09$fm)
 and one volume ($L\sim 2.9$fm). They did check the pion mass dependence by simulating at three different pion masses
 ranging from 410MeV to 700MeV. Their conclusion was that, $Z_c(3900)$
 is a threshold cusp that is due to multi-channel interaction effects, see Ref.~\cite{Ikeda:2016zwx}.
 In particular, the $\eta_c\rho$ channel
 and the $J/\psi\pi$ channel all interact strongly with the $DD^*$ channel.
 It would really be nice to check this result using a different approach. For example,
 in the particular channel of $Z_c(3900)$, one could carry out a coupled channel
 study using L\"uscher formalism, which is feasible if one pre-selects say only two
 or three most important channels, and see if a similar conclusion could be reached.

 One should keep in mind that these studies discussed above are
 still quite preliminary to draw any definite conclusions.
 In particular, usually only one volume at one lattice spacing have been utilized
 in these lattice searches and more systematic studies are very much welcome here.
 Needless to say that the nature of these exotic structures,
 whether it is a resonance or a bound states or
 even just multichannel effects, remains a challenging problem
 and hopefully lattice will provide us with more information
 in the future.

\subsection{Other exotic structures from the lattice}

 Recently, Lang {\em et al} have also studied the counterparts of the
 above mentioned $XYZ$ particles in the bottom sector~\cite{Lang:2016jpk}.
 Near threshold exotics also show up in the charm-strange and bottom-strange
 mesons. For example, there have been lattice studies on the $D_s$
 and $B_s$ mesons~\cite{Cichy:2016bci,Lang:2015hza}.
 At this conference, there have also been a few reports on multi-heavy
 tetraquark states, see e.g. Ref.~\cite{Peters:2016isf}.
 HAL QCD have also reported their new results in baryon-baryon scattering, please
 refer to the review of Savage~\cite{Savage:2016} and references therein for further details.

\section{Summary and outlook}

 Generally speaking, lattice studies of spectroscopy has entered
 the precision stage. To be more specific, one has to focus on different
 subfields in lattice spectroscopy. In this short review, I have gone
 over a number of these developments in the past year or so.

 In hadron spectroscopy involving only the light quarks,
 especially for those stable ones under strong interaction,
 it is clear that the field has already entered
 the precision era. The lattice studies in this field are rather systematic and
 precise, not only to percent level, but in some cases to per mil level.
 People have also started to consider not only QCD but also QED simultaneously.
 We could also perform studies close enough to the physical pion mass point
 so that comparisons with effective field theories can be carried out.
 In fact, the scope of precise lattice computations goes beyond just the light quark sector.
 For charmonium below the open charm threshold, things are also rather precise,
 see e.g HPQCD's results~\cite{Donald:2012ga,Dowdall:2012ab} on the hyperfine splittings.
 Generally speaking, for hadrons that are stable under strong interaction,
 rather good accuracy have been obtained.

 However, for hadrons that can decay under strong interactions,
 one in principle needs to study the scattering process of the decay products
 and this is where L\"uscher formalism has come into play. In recent years,
 a lot of progress has been gained in this direction,
 both theoretically and in practical simulations. As we see in
 subsection~\ref{subsec:pipi-scattering}, rather good accuracy has been obtained
 for light meson scattering. Therefore, for single-channel scattering of light mesons, also rather precise
 and systematic results can be obtained. As I showed you in this review, numerous computations
 have been performed on the rho resonance. We have seen from Wilson's talk
 that people have also been able to tackle multi-channel scattering problems
 within L\"uscher formalism.

 For particles involving heavy quarks, especially those beyond the threshold, one
 has to deal with the scattering of the relevant hadrons. The complication here is
 that usually this is typically a multi-channel situation and a brut-force treatment using
 the conventional L\"uscher method is complicated.
 However, within certain approximations, the progress in this field is also steady,
 but more studies are definitely required. Although in this review, I only focused on the
 charmed meson case, lattice computations in this direction will definitely have
 very important impact on the experiments that have been fast developing
 in recent years. It is also desirable to search for other equivalent or
 complementary methods that can handle the multi-channel scattering of
 multi-hadron systems.

\section*{Acknowledgements}

The author would like to thank the members of the China Lattice QCD Collaboration (CLQCD) for
their continued support over the years. All figures appearing in this contribution come from
the relevant arXiv sources that are also cited in the reference section.
The author would also like to thank the following people for
useful information and critical discussions they share with me:
S.~Aoki, G.~Bali, R.~Briceno, A.~Cox, S.~Durr, A.~Francis, D.~Guo, C.~Helmes,
L.~Leskovec, K.F.~Liu, L.~Liu, Z.~Liu, U.~Meissner, S.~Prelovsek, A.~Rusetzky,
G.~Schierholz, F.~Stokes, M.~Wagner, D.~Wilson,
C.~Urbach.

 This work is supported in part by the
 National Science Foundation of China (NSFC) under the project
 No.11335001,                                                             % Key funding NSFC
 and by the Deutsche Forschungsgemeinschaft (DFG)
 and the NSFC (No.11621131001)                                            %Sino-German project
 through funds provided to the Sino-Germen CRC 110 ``Symmetries and the
 Emergence of Structure in QCD''.
 It is also supported by Ministry of Science and Technology of
 China (MSTC) under
 973 project "Systematic studies on light hadron spectroscopy",
 No. 2015CB856702.                                                         % 973 project

%\bibliographystyle{apsrev4-1}
%\nocite{apsrev41Control}
%\bibliography{spectrum_review_CLiu,revbib_custom}


%\begin{thebibliography}{99}
%\bibitem{BMW08}
%S.~D\"urr {\it et al.}, Science {\bf 322}, 1224\ (2008), arXiv:0906.3599 [hep-lat].

%\end{thebibliography}

\end{document}